# Limiting Current on Periodic Electron Sheets in a Planar Diode

David Chernin[1], Dion Li[2,a], and Y.Y. Lau[2]

[1]Leidos, Inc., Reston, VA 20190

[2]Department of Nuclear Engineering and Radiological Sciences, University of Michigan, Ann Arbor, MI 48109

[a]Present address: Department of Physics, Massachusetts Institute of Technology, Cambridge, MA 02139

**ABSTRACT:** *We consider the steady state limiting current that can be carried by an infinite periodic array of thin electron sheets spaced by period p in a planar diode of gap voltage V and gap separation d. Our primary assumptions are (1) electron motion is restricted by an infinite magnetic field to the direction normal to the electrode surfaces, (2) all electrons are emitted from the cathode with initial kinetic energy $E_{in}$, and (3) electron motion is non-relativistic. The limiting current density, averaged over a period and normalized to the classical 1D Child-Langmuir (CL) current density (including a factor that accounts for non-zero $E_{in}$), is found to depend only on the two dimensionless parameters p/d and $E_{in}/eV$. This average limiting current density is computed from the maximum current density for which the iterative solution of a non-linear integral equation converges. Numerical results and empirical curve fits for the limiting current are presented, together with an analysis as p/d and $E_{in}/eV$ approach zero or infinity, in which cases previously published results are recovered. Our main finding is that, while the local anode current density within each electron sheet is infinite in our model (that is, it exceeds the classical 1D CL value by an 'infinite' factor), the period average anode current density is in fact still bounded by the classical 1D CL value. This study therefore provides further evidence that the classical 1D Child-Langmuir current density is truly a fundamental limit that cannot be circumvented.*

## I. INTRODUCTION

Cathode performance is an important technical issue for many applications [1],[2],[3]. It is very difficult to characterize because of emission non-uniformity [1],[4],[5],[6],[7],[8],[9], especially when there are actively emitting regions that are highly localized on the cathode surface. These localized emission hot spots, which may arise from regions of low work functions in a thermionic cathode for instance, could be the dominant contributors to the anode current even though they occupy a small fraction of the cathode area [7],[8]. Their presence prevents a routine characterization in terms of the classical, one-dimensional (1D) Child-Langmuir law (CLL) [10],[11], which specifies the maximum spatially uniform steady state current density that can flow between two infinite parallel plates separated by a distance $d$, to which a potential difference $V$ is applied,

$$J_{CL} = \frac{4\sqrt{2}}{9}\epsilon_0\sqrt{\frac{e}{m}}\frac{V^{3/2}}{d^2}, \tag{1.1}$$

where $-e$ and $m$ are the electron charge and mass, respectively, and $\epsilon_0$ is the free space permittivity. Many attempts have been made to generalize CLL to higher dimensions; an overwhelming majority failed when the active emission site has a scale very small compared with the anode-cathode spacing $d$.

Umstattd and Luginsland [12], Chernin et al. [7], and Jassem et al. [8] have shown that the limit (1.1) may be exceeded locally, near the edge of an emitting region, adjacent to a non-emitting region of the cathode surface. The reason is simply that the absence of the space charge along the non-emitting region means that additional charge must be present near the edge of the emitting region in order to reduce the surface electric field to zero, which is the current limiting condition, also known as the space-charge-limited condition. These papers [12],[7],[8] demonstrate the importance of highly localized emitting regions mentioned in the first paragraph. Taking this effect to its limit, the present authors [13] recently showed that it is even possible to exceed $J_{CL}$ locally by an 'infinite' factor, by showing that solutions exist for an isolated electron sheet of infinitesimal width, though the existence of such solutions requires that the electrons be emitted with a finite initial velocity. This discovery led to the natural question whether the *average* current density of a periodic array of such '$\delta$-function' sheets might exceed $J_{CL}$ (as modified to account for a finite emission velocity [14]). In the present paper we demonstrate that the answer to this question is 'no', thereby providing further evidence that $J_{CL}$ is truly a fundamental limit that cannot be exceeded.

This paper is organized as follows. Section II describes the model and the governing integral equation that determines the limiting current carried by periodic sheets of emitting electrons in a planar diode. Section III presents the numerical results obtained from this integral equation, together with an examination of the limiting cases inaccessible from numerical computation. Section IV concludes the study. The Appendices provide the mathematical details for Sections II and III, and an analytic fitting formula for the numerical data.

## II. FORMULATION

Our model consists of an infinite, periodic series of thin electron sheets spaced by period $p$ in a planar diode with gap separation $d$ and gap voltage $V$ (Fig. 1). An infinite magnetic field in the z-direction is assumed so that all electron motions are restricted to the z-direction. The electron sheets are infinitesimally thin. All electrons are assumed to be emitted from the cathode at z = 0 with the same energy $E_{in} = eV_{in} = (1/2)mv_{in}^2$. Since the solutions are periodic in $x$ with period $p$, we may focus on the single period, $-p/2 \leq x \leq p/2$, and on the electron sheet at $x = 0$. Since all quantities are independent of $y$, the electrostatic potential $\phi(z)$ on this electron sheet has only a z-dependence, and the velocity of an electron on this sheet is given by $v(z) = [(2/m)(E_{in} + e\phi(z))]^{1/2}$. The magnitude of the surface charge density is $\sigma(z) = M_2/v(z)$, where $M_2$ ($> 0$, in A/m) is a constant measuring the current carried by each electron sheet per unit length in $y$ in this 2-dimensional model (Fig. 1). We remark that $M_2$, and its corresponding dimensionless parameter $K_2$ given in Eq. (2.2) below, are the same as in [13] where an isolated, single electron sheet was considered (cf. Eq. (3.11) of [13]).

The electrostatic potential $\phi(z)$ consists of two components, the vacuum potential, $Vz/d$, and the potential due to the space charge on all electron sheets, which are implicitly included in the periodic solutions. The latter component is proportional to $M_2$, and is derived in Appendix A. This leads to the integral equation for $\phi(z)$ which, in terms of the dimensionless variables $\bar{\phi} = \phi/V$, $\bar{z} = z/d$, $\bar{z}_c = z_c/d$, reads

$$\bar{\phi}(\bar{z}) = \bar{z} - 2\pi K_2 \left\{ \int_0^{\bar{z}} \frac{\bar{G}_p(1-\bar{z},\bar{z}_c) d\bar{z}_c}{(\bar{\phi}(\bar{z}_c)+\Delta)^{1/2}} + \int_{\bar{z}}^1 \frac{\bar{G}_p(\bar{z},1-\bar{z}_c) d\bar{z}_c}{(\bar{\phi}(\bar{z}_c)+\Delta)^{1/2}} \right\}, \quad 0 \leq \bar{z} \leq 1, \tag{2.1}$$

$$K_2 = \frac{2}{9\pi} \times \frac{M_2}{dJ_{CL}} \equiv \frac{2}{9\pi} \times \left(\frac{p}{d}\right)\left(\frac{J_{av}}{J_{CL}}\right), \tag{2.2}$$

$$\Delta = E_{in}/eV, \tag{2.3}$$

where $K_2$ ($> 0$) is the dimensionless parameter measuring the sheet current, $J_{av} = M_2/p$ ($> 0$, in A/m²) is the average current density per period, $J_{CL}$ is the 1D classical Child-Langmuir current density, Eq. (1.1), and $\Delta$ is the dimensionless parameter measuring the injection energy of the mono-energetic electrons. The dimensionless Green's function in Eq. (2.1) is given by [cf. Appendix A],

$$\bar{G}_p(\bar{z}_1, \bar{z}_2) = \frac{1}{\bar{p}}[\bar{z}_1\bar{z}_2 + f(\bar{z}_1, \bar{z}_2)], \tag{2.4}$$

$$f(\bar{z}_1, \bar{z}_2) = 2\sum_{n=1}^{\infty} \frac{\sinh(\bar{k}_n\bar{z}_1)\sinh(\bar{k}_n\bar{z}_2)}{\bar{k}_n \sinh(\bar{k}_n)}, \tag{2.5}$$

$$\bar{p} = p/d, \quad \bar{k}_n = 2n\pi/\bar{p}. \tag{2.6}$$

The limiting current on an electron sheet of the periodic array is given by the maximum value of $K_2$ beyond which there is no solution to the integral equation (2.1). Note that $\bar{G}_p(\bar{z}_1, \bar{z}_2)$ depends only on $\bar{p}$, and that the maximum value of $K_2$, denoted as $K_2(\max)$, depends only on the two dimensionless parameters, $\Delta$ and $\bar{p}$. In Eq. (2.1), the first term ($\bar{z}$) represents the vacuum potential, $Vz/d$, and the second term, proportional to $K_2$ or $M_2$, represents the potential due to the space charge from all electron sheets (Fig. 1). We shall show that, in the limit $\bar{p} \to \infty$, Eq. (2.1) reduces to the integral equation for a single, isolated electron sheet that was treated in detail by Lau et al. [13].

As in [13], the integral equation (2.1) is solved iteratively for finite, nonzero values of $\Delta$ and $\bar{p}$, starting with the vacuum field solution, $\bar{\phi}(\bar{z}) = \bar{z}$. The approximate solution after the $k$-th iteration is given by,

$$\bar{\phi}^{(k)}(\bar{z}) = \bar{z} - 2\pi K_2 \left\{ \int_0^{\bar{z}} \frac{\bar{G}_p(1-\bar{z},\bar{z}_c) d\bar{z}_c}{(\bar{\phi}^{(k-1)}(\bar{z}_c)+\Delta)^{1/2}} + \int_{\bar{z}}^1 \frac{\bar{G}_p(\bar{z},1-\bar{z}_c) d\bar{z}_c}{(\bar{\phi}^{(k-1)}(\bar{z}_c)+\Delta)^{1/2}} \right\}, \quad k = 1,2,3,\ldots, \quad \bar{\phi}^{(0)}(\bar{z}) = \bar{z}. \tag{2.7}$$

Since the Green's function $\bar{G}_p$ is an infinite series which diverges logarithmically at $\bar{z}_c = \bar{z}$, the iterative solution for the maximum value of $K_2$ is computationally more demanding than the problem solved in [13]. The numerical algorithm to solve Eq. (2.7) iteratively is described toward the end of Appendix A.

In Section III, we present the numerical data on the maximum value, $K_2(\max)$, at various values of $\bar{p}$ and $\Delta$. In terms of $K_2(\max)$, the maximum period average current density, $J_{av}(\max)$, may be obtained from Eq. (2.2),

$$\frac{J_{av}(\text{max})}{J_{CL}} = \frac{9\pi}{2} \times \frac{K_2(\text{max})}{\bar{p}}. \tag{2.8}$$

This period average limiting current density is more conveniently compared with the classical 1D CL value modified by a nonzero Δ. With a nonzero Δ, Jaffe [14] modified Eq. (1.1) to read

$$J_{CL-J} = J_{CL} \times \left[(1+\Delta)^{1/2} + \Delta^{1/2}\right]^3. \tag{2.9}$$

The maximum period average current density normalized to $J_{CL-J}$, denoted by $\bar{J}_{max}$, becomes

$$\bar{J}_{max}(\Delta, \bar{p}) \equiv \frac{J_{av}(\text{max})}{J_{CL-L}} = \frac{9\pi}{2} \times \frac{K_2(\text{max})}{\bar{p}\left[(1+\Delta)^{1/2}+\Delta^{1/2}\right]^3}. \tag{2.10}$$

The value of $\bar{J}_{max}(\Delta, \bar{p})$ was determined using a simple bisection algorithm described in the last paragraph of Appendix A.

### III. LIMITING CURRENT ON PERIODIC ELECTRON SHEETS

Figure 2 shows $\bar{J}_{max}(\Delta, \bar{p})$ as a function of $\bar{p}$ for various values of Δ. Figure 3 shows $\bar{J}_{max}$ as a function of Δ for various values of $\bar{p}$. Data is obtained only for a limited range in Δ: $\Delta = 0.001, 0.01, 0.1, 2, 10$, and in $\bar{p}$: $\bar{p} = 0.05$ up to 3. The numerical fits for the data (Appendix E) within these ranges of Δ and $\bar{p}$ are shown by the dashed curves in Figs. 2 and 3. The analytic properties of $K_2(\text{max})$, in the limits of Δ and $\bar{p}$ approaching zero and infinity, are summarized in this section. The details are given in the Appendices.

Figures 2 and 3 reveal the following properties of $\bar{J}_{max}$.

(A) As $\bar{p} \to 0$, $\bar{J}_{max} \to 1$. This may be expected intuitively, because in this case, the periodic electron sheets are packed together infinitely closely, since $\bar{p} = p/d \to 0$. The average (or period-average) limiting current density should then approach the classical 1D Child-Langmuir law, corrected by Jaffe for nonzero Δ, Eq. (2.9). This is proven in Appendix B. Note that Eq. (2.10) yields the analytic result on $K_2(\text{max})$ for this case,

$$K_2(\text{max}) = \frac{2}{9\pi}\bar{p}\left[(1+\Delta)^{1/2} + \Delta^{1/2}\right]^3, \quad \bar{p} \to 0. \tag{3.1}$$

(B) As $\bar{p} \to \infty$, the sheet separation is infinite, and one electron sheet is unaffected by any of its neighbors (Fig. 1). The maximum current in this limit must be the same as that for a single, isolated electron sheet [13]. Appendix C shows that, as $\bar{p} \to \infty$, $K_2(\text{max})$ obtained from Eq. (2.1) indeed reduces to $K_2(\Delta)$ for a single, isolated electron sheet that is shown in Fig. 6 of [13]. Since $K_2(\Delta)$ is finite for all Δ, Eq. (2.10) gives the following expression for $\bar{J}_{max}$ in this single sheet limit,

$$\bar{J}_{max}(\Delta, \bar{p} \to \infty) = \frac{9\pi}{2} \times \frac{K_2(\Delta)}{\bar{p}\left[(1+\Delta)^{1/2}+\Delta^{1/2}\right]^3}. \tag{3.2}$$

When the period is infinite, the period average current density must be zero, as confirmed by Eq. (3.2), and suggested in Figs. 2 and 3. From Fig. 6 of [13], an excellent fitting formula for $K_2(\Delta)$ at small Δ reads,

$$K_2(\Delta) \cong 0.2336 \times \Delta^{0.5274}, \quad 0 < \Delta < 0.01. \tag{3.3}$$

(C) As $\Delta \to 0$, $\bar{J}_{max} \to 0$ for all nonzero values of $\bar{p}$. This trend is suggested in Figs. 2 and 3, and its validity may be demonstrated with the following argument. When $\bar{p}$ is finite and nonzero, the Green's function $\bar{G}_p$ that appears in both integrals in Eq. (2.1) may be shown to contain a logarithmic singularity at $\bar{z}_c = \bar{z}$. (This logarithmic singularity also appears in Eq. (B.6) in the $\bar{p} \to 0$ limit, and in Eq. (C.9) in the $\bar{p} \to \infty$ limit.) If $\Delta = 0$, this singularity at $\bar{z}_c = \bar{z}$ always leads to an arbitrarily large negative value for the curly bracket in Eq. (2.1) at some $\bar{z} \in (0,1)$, thus forcing a null value of $K_2$ as the only solution to Eq. (2.1). Such null solutions were examined in great detail, and properly interpreted in [13]. Comparing (A) with (C), one observes nonuniform convergence of $\bar{J}_{max}(\Delta, \bar{p})$ in the double limits, $\Delta = 0$ and $\bar{p} = 0$. It is this nonuniform convergence that led to the considerable difficulty in the numerical solution to the integral equation (2.7), especially in the limit $\Delta \to 0$.

(D) As $\Delta \to \infty$, $\bar{J}_{max}$ is independent of Δ at a fixed value of $\bar{p}$. This statement is proved in Appendix D. This explains why the $\Delta = 2$ and $\Delta = 10$ curves in Fig. 2 are almost indistinguishable, and why all curves in Fig. 3 become horizontal at large Δ. Note that the mathematical limit $\Delta \to \infty$ corresponds to the physically significant limit of a short circuit diode in Fig. 1, because in this case we may consider the gap voltage $V \to 0$ so that $\Delta = E_{in}/eV \to \infty$ at any fixed, nonzero injection energy $E_{in}$ of the electrons. The governing equation for, and the solution to $\bar{J}_{max}$ are described in Appendix D for this infinite Δ limit. We also point out in Appendix D that the curve $\bar{J}_{max}(\Delta = \infty, \bar{p})$ as a function of $\bar{p}$ is indistinguishable from the $\Delta = 2$ and $\Delta = 10$ curves in Fig. 2.

(E) Since $\bar{J}_{max}(\Delta, \bar{p})$ could not be computed readily from the integral equation (2.1), and Figs. 2 and 3 exhibit complex features, we include in Appendix E a formula that provides a ready-to-use analytical fit for $\bar{J}_{max}(\Delta, \bar{p})$ over the ranges of $\Delta$ and $\bar{p}$ shown: $\Delta = 0.001, 0.01, 0.1, 2, 10$, and $\bar{p} = 0.05$ up to 3. This analytical fit is included in Figs. 2 and 3 where excellent agreement is noted in its comparison with the numerical results from the integral equation. The analytical fit over this finite range, together with the asymptotic properties outlined in (A) – (D) above, could be useful for future design in 2D vacuum microelectronics.

## IV. CONCLUDING REMARKS

This paper shows that, while each electron sheet may carry a local current density that is infinitely large compared with the Child-Langmuir-Jaffe value, Eq. (2.9), the average current density of a periodic array of such sheets may approach this value, but never exceeds it. Likewise, in 2D and 3D simulations of thermionic cathodes even with highly localized active emission regions [7],[8],[12] the average anode current density may approach, but never exceed the classical CLL that includes a small thermal correction to Eq. (1.1) [10],[15]. This strongly suggests that the 1D classical CLL is a hard limit that cannot be exceeded, in thermal or non-thermal 2D or 3D models over vastly different forms and degrees of emission nonuniformity. This speculation applies regardless of the cathode's material properties, and is drawn from extensive analyses under the assumption of a smooth cathode surface.

When cathode surface roughness is present, local enhancement of the surface electric field could lead to strong local field emission of electrons. One might argue that such a strong local emission might produce additional, local hot spots, whose effects qualitatively resemble a modification of the local work function or local surface temperature on an otherwise flat emitting surface. Using this argument, one might venture that the average anode current density is bounded by the 1D CLL under steady state operation for all types of cathodes, whether they be thermionic, field emission, plasma-based or photo-cathodes. Note that the CLL may also be interpreted as a restriction on the total charge, Q ~ CV, imposed on a diode of vacuum capacitance C [16],[17],[18].

For pulsed operation, especially when the pulse length is less than the electron transit time across the diode, the instantaneous current density on the anode might exceed the 1D CCL, but the total charge Q within the diode is still found to be bounded by Q ~ CV, just like the steady state operation [19]. Thus, Q ~ CV appears to govern the maximum total charge within a diode in general, whether the electron emission is uniform in space or in time, and is independent of the emission mechanism or the conditions of the cathode surface.

## V. ACKNOWLEDGMENTS

This work was supported in part by the Air Force Office of Scientific Research (AFOSR) under Grant FA9550-20-1-0409, and Grant FA9550-21-1-0184. The work of D.L. is also supported by a graduate fellowship of the National Science Foundation.

## VI. DATA AVAILABILITY

The data that support the findings of this study are available from the corresponding author upon reasonable request.

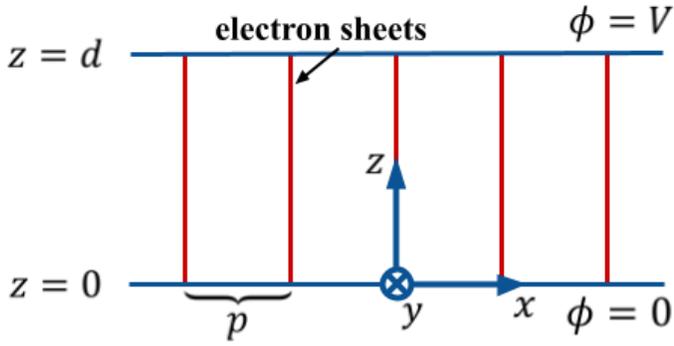

Fig. 1. A two-dimensional planar diode of gap spacing $d$ and gap voltage $V$. Electrons are emitted from the cathode ($z = 0$) in the z-direction with initial energy $E_{in}$, in the form of periodic electron sheets of separation $p$ and infinitesimal thickness (in red).

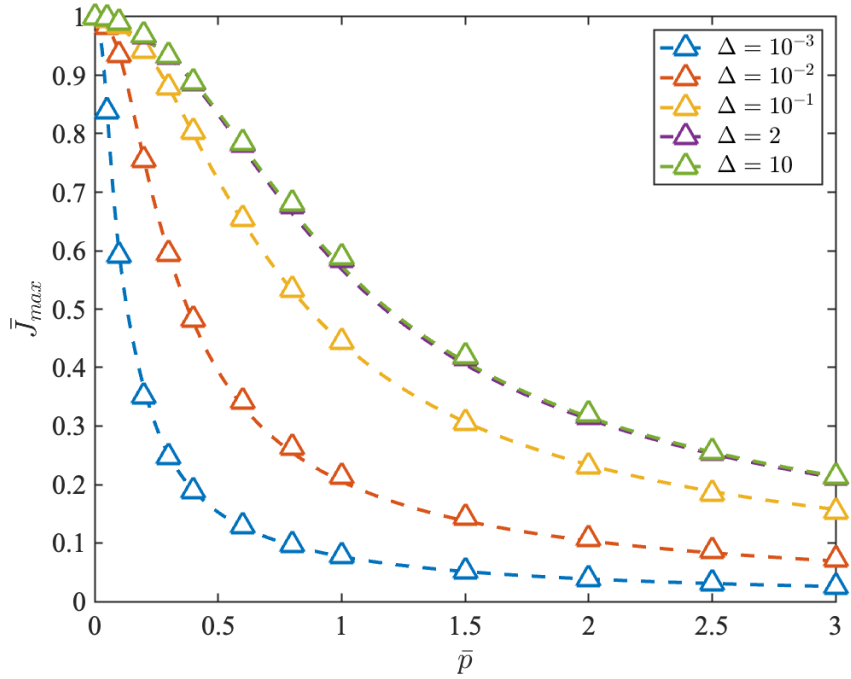

Fig. 2. The limiting current density $\bar{J}_{max}(\Delta, \bar{p})$, averaged over a period and normalized to the Child-Langmuir-Jaffe value Eq. (2.9), as a function of $\bar{p}$ for various values of $\Delta$. The triangles represent the solutions to the integral equation (2.1) and the dashed lines represent the numerical fit (see Appendix E). Note that the $\Delta = 2$ and $\Delta = 10$ curves are indistinguishable on the scale shown (see Appendix D).

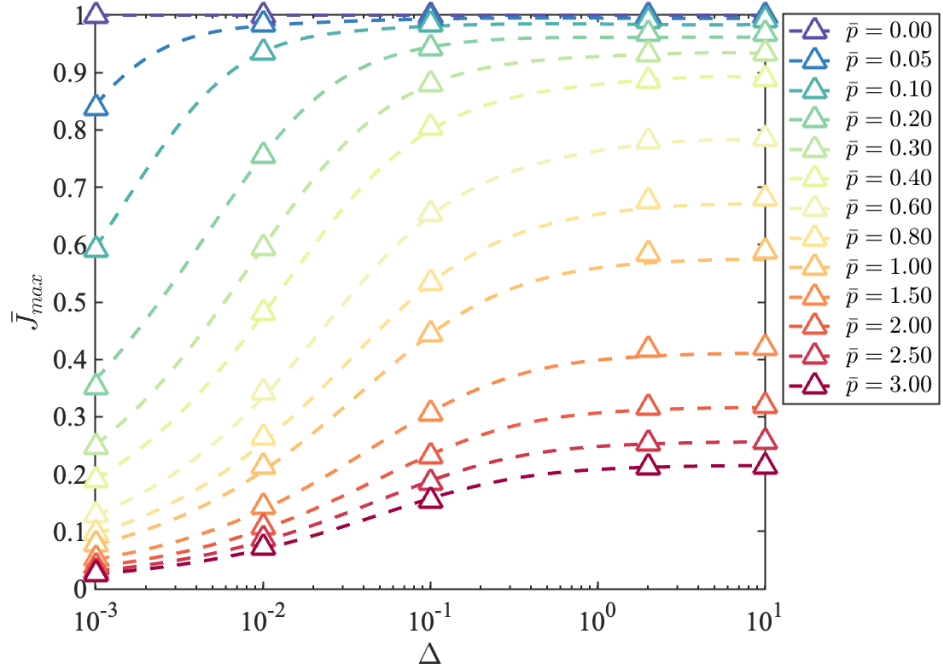

Fig. 3. The limiting current density $\bar{J}_{max}(\Delta, \bar{p})$, averaged over a period and normalized to the Child-Langmuir-Jaffe value Eq. (2.9), as a function of $\Delta$ for various values of $\bar{p}$. The triangles represent the solutions to the integral equation (2.1) and the dashed lines represent the numerical fit (see Appendix E).

## APPENDIX A. DERIVATION OF EQ. (2.1)

In this Appendix, we outline the derivation of Eq. (2.1) and summarize the numerical algorithm for its solution. In Fig. 1, the electrons are acted upon by the electric field produced by the combination of an applied potential difference $V$ between the plates and the space charge of all of the sheets. It suffices to consider a single period, $-p/2 \leq x \leq p/2$, $0 \leq z \leq d$, and the electron sheet at $x = 0$ (Fig. 1). All quantities are independent of $y$. We assume an infinite magnetic field in the z-direction, since it was established that the limiting current is insensitive to the imposed longitudinal magnetic field [7],[8],[12]. All electrons are emitted at $z = 0$ in the z-direction with kinetic energy $E_{in} = mv_{in}^2/2$. The current density $J(x, z) = M_2 \delta(x)$ where $M_2$ $(> 0)$ is a constant, independent of $z$, and $\delta$ is the Dirac delta function. Poisson's equation for the potential $\Phi(x, z)$ is then

$$\left(\frac{\partial^2}{\partial x^2} + \frac{\partial^2}{\partial z^2}\right)\Phi(x,z) = \frac{M_2}{\varepsilon_0 v_{in}} \frac{\delta(x)}{\left(1+\frac{e\phi(z)}{E_{in}}\right)^{1/2}} \equiv S(z)\delta(x) \qquad (A.1)$$

where we have defined the source strength $S(z)$ and where $\phi(z) \equiv \Phi(0, z)$ is the potential encountered by the electrons within the sheet. We require the potential $\Phi(x, z)$ to satisfy the boundary conditions $\Phi(x, 0) = 0$, $\Phi(x, d) = V$ and $\Phi(-p/2, z) = \Phi(+p/2, z)$. Our goal is to find an equation for $\phi(z)$.

We begin by expanding the potential in a cosine series in $x$,

$$\Phi(x, z) = \frac{1}{2}\Phi_0(z) + \sum_{n=1}^{\infty} \Phi_n(z) \cos(k_n x) \qquad (A.2)$$

where $k_n = 2\pi n/p$ and the coefficient functions $\Phi_n(z)$ are given by

$$\Phi_n(z) = \frac{2}{p} \int_{-p/2}^{p/2} dx\, \Phi(x, z) \cos(k_n x). \qquad (A.3)$$

Using the formal expansion of the delta function,

$$\delta(x) = \frac{1}{p} + \frac{2}{p}\sum_{n=1}^{\infty} \cos(k_n x) \qquad (A.4)$$

it follows from Eq. (A.1) that the coefficient functions must satisfy

$$\left(\frac{d^2}{dz^2} - k_n^2\right)\Phi_n(z) = \frac{2}{p}S(z) \tag{A.5}$$

along with the boundary conditions $\Phi_n(0) = 0$ and $\Phi_n(d) = 2V\delta_{n,0}$, where $\delta_{n,m}$ is the Kronecker delta. The solution of Eq. (A.5) subject to the given boundary conditions is elementary. Substituting that solution in Eq. (A.2) and setting $x = 0$ gives a non-linear integral equation for $\phi(z)$:

$$\phi(z) = Vz/d - \int_0^z dz_c G(d-z, z_c) S(z_c) - \int_z^d dz_c G(z, d-z_c) S(z_c) \tag{A.7}$$

which is Eq. (2.1) in the main text, where

$$G(z_1, z_2) \equiv \frac{z_1 z_2}{pd} + 2\sum_{n=1}^{\infty} \frac{\sinh(k_n z_1)\sinh(k_n z_2)}{(k_n p)\sinh(k_n d)}. \tag{A.8}$$

The sum in Eq. (A.8) converges when $z_1 + z_2 < d$, but diverges logarithmically when $z_1 + z_2 = d$. Consequently, the integrands in Eq. (A.7) have logarithmic singularities at the endpoints $z_c = z$, which require careful numerical treatment. See [13] for a discussion of this logarithmic singularity, which also occurs below in Eq. (B.6) and Eq. (C.10).

**Numerical algorithms for the iterative solutions**

Equation (2.1) is solved iteratively using Eq. (2.7). The iteration proceeds until one of three things happen: (1) The fractional difference between $\bar{\phi}^{(k)}(\bar{z})$ and $\bar{\phi}^{(k-1)}(\bar{z})$ is less than a specified maximum for all points on the z-grid, (2) The argument of the square root in Eq. (2.7) becomes negative at any grid point, or (3) A maximum number of iterations is reached. The iteration is considered to be converged if and only if (1) is satisfied. For the numerical results shown in Figs. 2 and 3, we used $10^{-6}$ for the maximum fractional difference in the convergence condition (1) and 200 for the maximum number of iterations in (3). All calculations used 20,000 steps in the interval $\bar{z} = [0,1]$ to evaluate the integrals. This large number is required to resolve the potential minimum for small values of $\Delta$. The integrals were evaluated using the "midpoint" method, described in Appendix B of [13]. The sum in Eq. (2.5) was truncated when the n-th term was less than $10^{-6}$ times the previously accumulated sum.

The value of $\bar{J}_{max}(\Delta, \bar{p})$ in Eq. (2.10) was determined using a simple bisection algorithm, as follows: For an assigned pair of nonzero value $(\Delta, \bar{p})$, we start with values of $\bar{J}$ that bracket the expected value of $\bar{J}_{max}$; in particular, we start with $\bar{J}_1 = 0$ and $\bar{J}_2 = 2$, such that we anticipate that the iteration (2.7) will converge for $\bar{J} = \bar{J}_1$ and will not converge for $\bar{J} = \bar{J}_2$. We then try the iteration (2.7) for a value of $\bar{J}_{mid} \equiv (\bar{J}_1 + \bar{J}_2)/2$. If this iteration converges, we assign a new value for $\bar{J}_1 = \bar{J}_{mid}$; if the iteration does not converge, we assign a new value for $\bar{J}_2 = \bar{J}_{mid}$. We repeat this bisection 12 times, which gives us $\bar{J}_{max}$ to an accuracy of $2/2^{12}$, or approximately $5 \times 10^{-4}$.

## APPENDIX B. THE LIMIT $\bar{p} \to 0$

We first show that in the limit $\bar{p} \to 0$ the term $f(\bar{z}_1, \bar{z}_2)$ in Eq. (2.4) contributes negligibly to both integrals in Eq. (2.1), in comparison with its preceding term, $\bar{z}_1 \bar{z}_2$. As $\bar{p} \to 0$, all "sinh" terms in Eq. (2.5) are exponentially large. We may thus write,

$$f(\bar{z}_1, \bar{z}_2) \sim \sum_{n=1}^{\infty} \frac{1}{(2n\pi/\bar{p})} e^{-(2n\pi/\bar{p}) + (2n\pi/\bar{p})(\bar{z}_1 + \bar{z}_2)}. \tag{B.1}$$

Using Eq. (2.4) in Eq. (2.1), we see that $\bar{z}_1 + \bar{z}_2 = 1 - (\bar{z} - \bar{z}_c)$ for the first integral in Eq. (2.1), and that $\bar{z}_1 + \bar{z}_2 = 1 + (\bar{z} - \bar{z}_c)$ for the second integral in Eq. (2.1). For both integrals, we may write in the compact form,

$$\bar{z}_1 + \bar{z}_2 = 1 - |\bar{z} - \bar{z}_c|, \tag{B.2}$$

and Eq. (B.1) becomes,

$$f(\bar{z}_1, \bar{z}_2) = \frac{\bar{p}}{2\pi} g(\xi), \quad (\bar{p} \to 0) \tag{B.3}$$

$$g(\xi) = \sum_{n=1}^{\infty} \frac{1}{n} e^{-n\xi}, \quad \xi = 2\pi|\bar{z} - \bar{z}_c|/\bar{p}. \tag{B.4}$$

In Eq. (B.4), the infinite sum for $g(\xi)$ converges for all $\xi$ except at $\xi = 0$. As $\xi \to 0$, we approximate $dg(\xi)/d\xi = -\sum_{n=1}^{\infty} e^{-n\xi} = -e^{-\xi}/(1 - e^{-\xi}) \cong -1/\xi$. Integrating and using Eq. (B.4), we obtain,

$$g(\xi) \cong -\ln|\xi| = -\ln(2\pi) + \ln(\bar{p}) - \ln|\bar{z} - \bar{z}_c|, \quad (\xi \to 0). \tag{B.5}$$

Substitute Eq. (B.5) into Eq. (B.3) to obtain,

$$f(\bar{z}_1, \bar{z}_2) \cong \frac{\bar{p}}{2\pi}[-\ln(2\pi) + \ln(\bar{p}) - \ln|\bar{z}-\bar{z}_c|], \quad (\bar{z} \to \bar{z}_c, \bar{p} \to 0). \tag{B.6}$$

Note that the logarithmic singularity at $\bar{z}_c = \bar{z}$ in Eq. (B.6) is integrable in both integrals in Eq. (2.1), upon using Eq. (2.4). Equations (B.6) and (B.3) thus show that, as $\bar{p} \to 0$, the contribution from $f(\bar{z}_1, \bar{z}_2)$ is negligible compared with the first term in Eq. (2.4) and we may approximate

$$\bar{G}_p(\bar{z}_1, \bar{z}_2) \simeq \frac{\bar{z}_1 \bar{z}_2}{\bar{p}}, \tag{B.7}$$

for both integrals in Eq. (2.1). Upon substituting Eq. (B.7) into Eq. (2.1), Eq. (2.1) is identical to Eq. (3.4) of Lau et al. [13], in which the parameter $K_1 = 2\pi K_2/\bar{p} = (4/9)J/J_{CL}$ may be identified. Including Jaffe's correction for nonzero $\Delta$, Eq. (2.9), this means $K_2 = K_2(\max)$, where

$$K_2(\max) = \frac{2}{9\pi}\bar{p}\left[(1+\Delta)^{1/2} + \Delta^{1/2}\right]^3, \quad \bar{p} \to 0, \tag{B.8}$$

which is Eq. (3.1).

## APPENDIX C. THE LIMIT $\bar{p} \to \infty$

As $\bar{p} \to \infty$, the first term in Eq. (2.4) vanishes, leaving behind the second term,

$$\bar{G}_p(\bar{z}_1, \bar{z}_2) = \frac{f(\bar{z}_1, \bar{z}_2)}{\bar{p}} = \frac{2}{\bar{p}} \sum_{n=1}^{\infty} \frac{\sinh(\bar{k}_n \bar{z}_1)\sinh(\bar{k}_n \bar{z}_2)}{\bar{k}_n \sinh(\bar{k}_n)} \equiv \bar{G}_\infty(\bar{z}_1, \bar{z}_2). \tag{C.1}$$

Since $\bar{k}_n = 2n\pi/\bar{p}$, the infinite sum in Eq. (C.1) may be converted into an integral as $\bar{p} \to \infty$, with the substitutions,

$$\bar{k}_n \to k, \ 1/\bar{p} \to dk/2\pi, \ \sum_{n=1}^{\infty} \to (\bar{p}/2\pi)\int_0^\infty dk, \tag{C.2}$$

$$\bar{G}_\infty(\bar{z}_1, \bar{z}_2) = \frac{1}{\pi}\int_0^\infty \frac{dk}{k}\frac{\sinh(k\bar{z}_1)\sinh(k\bar{z}_2)}{\sinh(k)}. \tag{C.3}$$

We next differentiate Eq. (C.3) with respect to $\bar{z}_1$ to obtain,

$$\partial \bar{G}_\infty/\partial \bar{z}_1 = \frac{1}{\pi}\int_0^\infty dk \frac{\cosh(k\bar{z}_1)\sinh(k\bar{z}_2)}{\sinh(k)} = \frac{1}{4}\left[\tan\left(\frac{\pi}{2}(\bar{z}_1+\bar{z}_2)\right) - \tan\left(\frac{\pi}{2}(\bar{z}_1-\bar{z}_2)\right)\right], \tag{C.4}$$

where we have used the identity, $\cosh(x)\sinh(y) = [\sinh(x+y) - \sinh(x-y)]/2$, and

$$\int_0^\infty dk \frac{\sinh(kz)}{\sinh(k)} = \frac{\pi}{2}\tan\left(\frac{\pi}{2}z\right), \tag{C.5}$$

to derive the last expression of Eq. (C.4). Integrating Eq. (C.4) with respect to $\bar{z}_1$, we have

$$\bar{G}_\infty(\bar{z}_1, \bar{z}_2) = -\frac{1}{2\pi}\ln\left|\frac{\cos\left(\frac{\pi}{2}(\bar{z}_1+\bar{z}_2)\right)}{\cos\left(\frac{\pi}{2}(\bar{z}_1-\bar{z}_2)\right)}\right|, \tag{C.8}$$

which yields

$$\bar{G}_p(1-\bar{z}, \bar{z}_c) = \bar{G}_p(\bar{z}, 1-\bar{z}_c) = -\frac{1}{2\pi}\ln\left[\frac{|\sin(\frac{\pi}{2}(-\bar{z}+\bar{z}_c))|}{\sin(\frac{\pi}{2}(\bar{z}+\bar{z}_c))}\right], \quad \bar{p} \to \infty. \tag{C.9}$$

Substitution of Eq. (C.9) into Eq. (2.1) yields,

$$\bar{\phi}(\bar{z}) = \bar{z} + K_2 \int_0^1 \frac{d\bar{z}_c}{(\bar{\phi}(\bar{z}_c)+\Delta)^{1/2}} \ln\left[\frac{|\sin(\frac{\pi}{2}(-\bar{z}+\bar{z}_c))|}{\sin(\frac{\pi}{2}(\bar{z}+\bar{z}_c))}\right], \quad 0 \le \bar{z} \le 1, \ (\bar{p} \to \infty), \tag{C.10}$$

which is identical to Eq. (3.10) of [13], the integral equation for an isolated, single electron sheet whose normalized limiting current, $K_2$, is shown in Fig. 6 of [13] as a function of $\Delta$. This curve gives $K_2(\Delta)$, which is thus the same as $K_2(\max)$ in Eq. (2.10) in the limit $\bar{p} \to \infty$, yielding Eqs. (3.2) and (3.3) of the main text.

## APPENDIX D. THE LIMIT $\Delta \to \infty$

The limit $\Delta = E_{in}/eV \to \infty$ may either be treated as letting $E_{in} \to \infty$ at a fixed nonzero $V$, or letting $V \to 0$ at a fixed nonzero $E_{in}$. We find it more convenient to treat the $V \to 0$ limit at a finite, nonzero value of $E_{in}$. Note that the mathematical limit $V \to 0$ corresponds to a short circuit diode physically (Fig. 1). When $V \to 0$, the vacuum potential vanishes, and only the space charge potential remains. That is, the first term, $\bar{z}$, in Eq. (2.1) can be dropped, since it originates from the vacuum potential, $Vz/d$. Keeping all other terms, and defining $\bar{\psi} \equiv e\phi/E_{in} = \bar{\phi}/\Delta$, Eq. (2.1) reads,

$$\bar{\psi}(\bar{z}) = -K_\Delta \left\{ \int_0^{\bar{z}} \frac{\bar{G}_p(1-\bar{z},\bar{z}_c)d\bar{z}_c}{\left(\bar{\psi}(\bar{z}_c)+1\right)^{1/2}} + \int_{\bar{z}}^1 \frac{\bar{G}_p(\bar{z},1-\bar{z}_c)d\bar{z}_c}{\left(\bar{\psi}(\bar{z}_c)+1\right)^{1/2}} \right\}, \quad 0 \le \bar{z} \le 1, \tag{D.1}$$

where $K_\Delta$ ($> 0$) measures the normalized sheet current in this limit, $\Delta \to \infty$. The limiting current is determined by the maximum value of $K_\Delta$ beyond which there is no solution to the integral equation (D.1). Since $\bar{G}_p(\bar{z}_1, \bar{z}_2)$ depends only on $\bar{p}$, this maximum value of $K_\Delta$, denoted as $K_{\Delta\max}(\bar{p})$, is a function of $\bar{p}$ alone. Comparing Eq. (D.1) and Eq. (2.1), Eq. (2.10) yields,

$$\bar{J}_{max}(\Delta = \infty, \bar{p}) = \frac{9}{32} \times \frac{K_{\Delta\max}(\bar{p})}{\bar{p}}. \tag{D.2}$$

We have solved the integral equation (D.1) iteratively to obtain $K_{\Delta\max}(\bar{p})$, similar to Eq. (2.7), but starting with the vacuum field solution, $\bar{\psi}(\bar{z}) = 0$. This null vacuum solution followed from Eq. (D.1) with $K_\Delta = 0$, clearly expected when the gap voltage $V = 0$.

We find that the data points obtained from the iterative solution of Eq. (D.1) are indistinguishable from the data points for $\Delta = 10$ (and for $\Delta = 2$) in Fig. 2, for all nonzero values of $\bar{p}$.

## APPENDIX E. FITTING FORMULAS FOR $\bar{J}_{max}(\Delta, \bar{p})$

The numerical results for $\bar{J}_{max}(\Delta, \bar{p})$ obtained from the solutions of the integral equation (2.1) may be approximated by the following fitting expressions, over the ranges of $\Delta$ and $\bar{p}$ shown in Figs. 2 and 3,

$$\bar{J}_{fit}(\Delta, \bar{p}) = \tanh\left(\frac{1}{\bar{p}\alpha(\Delta)}\right) + \gamma(\Delta)(-1 + \cos(\beta(\Delta)\bar{p}))e^{-\beta(\Delta)\bar{p}}, \quad 0.001 < \Delta < 10, \quad 0.05 < \bar{p} < 3. \tag{E.1}$$

Here, $\alpha(\Delta), \beta(\Delta)$, and $\gamma(\Delta)$ can all be approximated using the following fitting model

$$\eta_{fit}(\Delta) = n_1(\ln \Delta)^4 + n_2(\ln \Delta)^3 + n_3(\ln \Delta)^2 + n_4 \ln \Delta + n_5, \quad 0.001 < \Delta < 10, \tag{E.2}$$

where $\eta$ represents, separately, $\alpha, \beta$, and $\gamma$. The $n_i$ ($i = 1, 2, 3, 4, 5$) values in Eq. (E.2) are given in Table 1. The fitting formula, Eq. (E.1), is shown by the dashed curves in Figs. 2 and 3. The deviation between Eq. (E.1) and the data points obtained from the integral equations is within 0.0369 percent. The fitting formula was obtained using MATLAB's Curve Fitting Toolbox. The sum of squares due to error (SSE) between Eq. (E.1) and the data points obtained from the integral equations are $1.292 \times 10^{-29}$, $5.443 \times 10^{-29}$, and $1.387 \times 10^{-32}$ for $\alpha, \beta$, and $\gamma$, respectively, and each fit had R-squared values very close to 1.

| coefficients\$\eta$ | $\alpha$ | $\beta$ | $\gamma$ |
|---|---|---|---|
| $n_1$ | 0.002393 | -0.008358 | 0.000367 |
| $n_2$ | -0.012280 | -0.102000 | 0.001815 |
| $n_3$ | 0.031840 | 0.169800 | -0.003380 |
| $n_4$ | -0.058670 | 0.389000 | -0.009625 |
| $n_5$ | 1.579000 | 4.985000 | 0.207600 |

Table 1. Values of the fitting coefficients $n_i$ ($i = 1, 2, 3, 4, 5$) for $\alpha(\Delta), \beta(\Delta)$, and $\gamma(\Delta)$.

# REFERENCES

[1]  A. S. Gilmour, *Klystrons, Traveling Wave Tubes, Magnetrons, Crossed-field Amplifiers, and Gyrotrons*. (Artech House, Norwood, MA, 2011).

[2]  R. J. Umstattd, D. Abe, J. Benford, D. Gallagher, R. M. Gilgenbach, D. M. Goebel, M. S. Litz, and J. A. Nation, "Cathodes and Electron Guns," in *High-Power Microwave Sources and Technologies*, Eds.: R. J. Barker and E. Schamiloglu, Ch. 9, pp. 284-324 (IEEE Press, Piscataway, NJ, 2001).

[3]  K. L. Jensen, *Introduction to the Physics of Electron Emission* (Wiley, Hoboken, NJ, 2017).

[4]  M. J. Cattelino, G. V. Miram, and W. R. Ayers, "A diagnostic technique for evaluation of cathode emission performance and defects in vehicle assembly," in *International Electron Devices Meeting*, San Francisco, CA, 1982.

[5]  E. A. Adler and R. T. Longo, "Effect of nonuniform work function on space-charge-limited current," *J. Appl. Phys.* **59**(4), 1022- 1027 (1986).

[6]  R. Vaughan, "A synthesis of the Longo and Eng cathode emission models," *IEEE Trans. Electron Devices* **33**(11), 1925–1927 (1986).

[7]  D. Chernin, Y. Y. Lau, J. J. Petillo, S. Ovtchinnikov, D. Chen, A. Jassem, R. Jacobs, D. Morgan, and J. H. Booske, "Effect of nonuniform emission on Miram curves," *IEEE Trans. Plasma Sci.* **48**(1), 146–155 (2020).

[8]  A. Jassem, D. Chernin, J. J. Petillo, Y. Y. Lau, A. Jensen, and S. Ovtchinnikov, "Analysis of anode current from a thermionic cathode with a 2-D work function distribution," *IEEE Trans. Plasma Sci.* **49**(2), 749–755 (2021).

[9]  D. Chen, R. Jacobs, D. Morgan, and J. Booske, "Physical factors governing the shape of the Miram curve knee in thermionic emission," *IEEE Trans. Electron Devices* **70**(3), 1219–1225 (2023).

[10]  C. D. Child, "Discharge From Hot CaO," *Phys. Rev. Ser. I*, 32 (5), 492–511 (1911).

[11]  I. Langmuir, "The effect of space charge and initial velocities on the potential distribution and thermionic current between parallel plane electrodes," *Phys. Rev.* **21**(4), 419–435 (1923).

[12]  R. J. Umstattd and J. W. Luginsland, "Two-dimensional space-charge-limited emission: beam-edge characteristics and applications," *Phys. Rev. Lett.* **87**(14), 145002 (2001).

[13]  Y. Y. Lau, D. Li, and D. P. Chernin, "On the Child-Langmuir law in one, two, and three dimensions," *Phys. Plasmas* **30**(9), 093104 (2023).

[14]  G. Jaffé, "On the currents carried by electrons of uniform initial velocity," *Phys. Rev.* **65**(3–4), 91–98 (1944).

[15]  T. C. Fry, "The thermionic current between parallel plane electrodes; velocities of emission distributed according to Maxwell's law," *Phys. Rev.* **17**(4), 441–452 (1921).

[16]  R. J. Umstattd, C. G. Carr, C. L. Frenzen, J. W. Luginsland, and Y. Y. Lau, "A simple physical derivation of Child–Langmuir space-charge-limited emission using vacuum capacitance," *Am. J. Phys.* **73**(2), 160–163 (2005).

[17]  P. Zhang, Á. Valfells, L. K. Ang, J. W. Luginsland, and Y. Y. Lau, "100 years of the physics of diodes," *Appl. Phys. Rev.* **4**(1), 011304 (2017).

[18]  Y. Y. Lau, J. Krall, M. Friedman, and V. Sirlin, "On certain theoretical aspects of relativistic klystron amplifiers," in *Microwave and Particle Beam Sources and Directed Energy Concepts*, SPIE Vol. 1061, 48-59 (1989).

[19]  Á. Valfells, D. W. Feldman, M. Virgo, P. G. O'Shea, and Y. Y. Lau, "Effects of pulse-length and emitter area on virtual cathode formation in electron guns," *Phys. Plasmas* **9**(5), 2377-2382 (2002).